\def\isarxiv{1} 

\ifdefined\isarxiv
\documentclass[11pt]{article}

\usepackage[numbers]{natbib}

\else
\documentclass{article}
\usepackage{neurips_2022}
\fi

\usepackage{amsmath}
\usepackage{amsthm}
\usepackage{amssymb}
\usepackage{algorithm}
\usepackage{subfig}
\usepackage{algpseudocode}
\usepackage{graphicx}
\usepackage{grffile}
\usepackage{wrapfig,epsfig}
\usepackage{url}
\usepackage{xcolor}
\usepackage{epstopdf}

\usepackage{bbm}
\usepackage{dsfont}

\allowdisplaybreaks

\ifdefined\isarxiv

\usepackage{tikz}
\usepackage{hyperref}  
\hypersetup{colorlinks=true,citecolor=blue,linkcolor=blue} 
\usetikzlibrary{arrows}
\usepackage[margin=1in]{geometry}

\else

\usepackage{microtype}
\usepackage{hyperref}
\definecolor{mydarkblue}{rgb}{0,0.08,0.45}
\hypersetup{colorlinks=true, citecolor=mydarkblue,linkcolor=mydarkblue}

\fi

\newtheorem{theorem}{Theorem}[section]
\newtheorem{lemma}[theorem]{Lemma}
\newtheorem{definition}[theorem]{Definition}

\newtheorem{fact}[theorem]{Fact}
\newtheorem{remark}[theorem]{Remark}
\newtheorem{claim}[theorem]{Claim}

\newcommand{\wh}{\widehat}
\newcommand{\wt}{\widetilde}

\newcommand{\R}{\mathbb{R}}

\newcommand{\Tmat}{{\cal T}_{\mathrm{mat}}}

\DeclareMathOperator*{\Z}{\mathbb{Z}}

\DeclareMathOperator{\poly}{poly}

\DeclareMathOperator{\SE}{\mathsf{SE}}
\DeclareMathOperator{\FAMP}{\mathsf{FAMP}}
\DeclareMathOperator{\SAMP}{\mathsf{SAMP}}

\makeatletter
\newcommand*{\RN}[1]{\expandafter\@slowromancap\romannumeral #1@}
\makeatother

\usepackage{lineno}

\begin{document}

\ifdefined\isarxiv

\date{}

\title{Revisiting Quantum Algorithms for Linear Regressions: Quadratic Speedups without Data-Dependent Parameters}
\author{
Zhao Song\thanks{\texttt{zsong@adobe.com}. Adobe Research.}
\and
Junze Yin\thanks{\texttt{junze@bu.edu}. Boston University.}
\and
Ruizhe Zhang\thanks{\texttt{ruizhe@utexas.edu}. Simons Institute at Berkeley.}
}

\else

\title{Intern Project} 
\maketitle 
\fi

\ifdefined\isarxiv
\begin{titlepage}
  \maketitle
  \begin{abstract}

Linear regression is one of the most fundamental linear algebra problems. Given a dense matrix $A \in \mathbb{R}^{n \times d}$ and a vector $b$, the goal is to find $x'$ such that 
 $   \| Ax' - b \|_2^2 \leq (1+\epsilon) \min_{x} \| A x - b \|_2^2 $.  
The best classical algorithm takes $O(nd) + \mathrm{poly}(d/\epsilon)$ time [Clarkson and Woodruff STOC 2013, Nelson and Nguyen FOCS 2013]. On the other hand, quantum linear regression algorithms can achieve exponential quantum speedups, as shown in [Wang Phys. Rev. A 96, 012335, Kerenidis and Prakash ITCS 2017, Chakraborty, Gily{\'e}n and Jeffery ICALP 2019]. However, the running times of these algorithms depend on some quantum linear algebra-related parameters, such as $\kappa(A)$, the condition number of $A$. In this work, we develop a quantum algorithm that runs in $\widetilde{O}(\epsilon^{-1}\sqrt{n}d^{1.5}) + \mathrm{poly}(d/\epsilon)$ time. It provides a quadratic quantum speedup in $n$ over the classical lower bound without any dependence on data-dependent parameters. In addition, we also show our result can be generalized to multiple regression and ridge linear regression.

  \end{abstract}
  \thispagestyle{empty}
\end{titlepage}

{\hypersetup{linkcolor=black}
\tableofcontents
}
\newpage

\else

\begin{abstract}

Linear regression is one of the most fundamental linear algebra problems. Given a dense matrix $A \in \mathbb{R}^{n \times d}$ and a vector $b$, the goal is to find $x'$ such that 
 $   \| Ax' - b \|_2^2 \leq (1+\epsilon) \min_{x} \| A x - b \|_2^2 $.  
The best classical algorithm takes $O(nd) + \mathrm{poly}(d/\epsilon)$ time [Clarkson and Woodruff STOC 2013, Nelson and Nguyen FOCS 2013]. On the other hand, quantum linear regression algorithms can achieve exponential quantum speedups, as shown in [Wang Phys. Rev. A 96, 012335, Kerenidis and Prakash ITCS 2017, Chakraborty, Gily{\'e}n and Jeffery ICALP 2019]. However, the running times of these algorithms depend on some quantum linear algebra-related parameters, such as $\kappa(A)$, the condition number of $A$. In this work, we develop a quantum algorithm that runs in $\widetilde{O}(\epsilon^{-1}\sqrt{n}d^{1.5}) + \mathrm{poly}(d/\epsilon)$ time. It provides a quadratic quantum speedup in $n$ over the classical lower bound without any dependence on data-dependent parameters. In addition, we also show our result can be generalized to multiple regression and ridge linear regression.

\end{abstract}

\fi

\section{Introduction}

Linear regression is one of the fundamental problems in machine learning and data science \cite{h67,ys09,f09}. It is a statistical method that models the relationship between a dependent variable and one or more independent variables. Multiple regression is an extension of linear regression that predicts a target variable using multiple feature variables. They have many applications across different areas such as predictive analysis \cite{kn19,vng09}, economics \cite{nkrs20,pls08,ajr01}, marketing \cite{bn98}, finance \cite{ggw23}, healthcare \cite{lj21,kkc+19,vrh+19,ta13}, education \cite{rs15,br99,omh20}, social sciences \cite{ug13,y23}, sports analytics \cite{st20,cw19}, manufacturing \cite{cb19,bm21}, and quality control \cite{qb12}.
 
The formal definition of linear regression can be described as follows:
\begin{definition}[Linear Regression]
Given a matrix $A \in \R^{n \times d}$ and a vector $b \in \R^{n}$, we let $\epsilon \in (0,1)$ denote an accuracy parameter. The goal is to output a vector $x \in \R^{d}$ such that
\begin{align*}
    \| A x - b \|_2^2 \leq (1+\epsilon) \min_{x' \in \R^{d }}  \| A x' - b \|_2^2
\end{align*}
\end{definition}
The state-of-the-art algorithms for solving linear regression are due to \cite{cw13,nn13}, where the running time is $O(nd) + \poly(d/\epsilon)$.

The formal definition of multiple regression can be described as follows:
\begin{definition}[Multiple Regression]

Given two matrices $A \in \R^{n \times d}$ and $B \in \R^{n \times N}$, we let $\epsilon \in (0,1)$ denote an accuracy parameter. The goal is to output a matrix $X \in \R^{d \times N}$ such that
\begin{align*}
    \| A X - B\|_F \leq (1+\epsilon)\min_{X' \in \R^{d \times N}} \| A X' - B \|_F
\end{align*}
    
\end{definition}

Ridge regression is a regularized version of linear regression that adds an $\ell_2$ penalty to the regression coefficients, preventing overfitting. This property makes it well-suited for handling high-dimensional data, where feature collinearity is common. In machine learning, ridge regression serves as a common baseline and benchmark method. Several studies have analyzed ridge regression concerning high-dimensional data and models \cite{dw18,m11}, feature selection \cite{pd16,zchd18,c08}, and regularization path algorithms \cite{fht10}. Moreover, it has found extensive use in diverse applications such as image recognition \cite{alv07,xzc09}, natural language processing \cite{bsk06,l21}, and bioinformatics \cite{xlll20,cvd11,bsk06}.

\begin{definition}[Ridge Regression]
    Given a matrix $A \in \R^{n \times d}$ and a vector $b \in \R^{n}$, we let $\epsilon \in (0,1)$ denote an accuracy parameter and let $\lambda > 0$ denote a regularization parameter. The goal is to output a vector $x \in \R^{d}$ such that
\begin{align*}
    \| A x - b \|_2^2 + \lambda \| x \|_2^2 \leq (1+\epsilon)\min_{x' \in \R^d} ( \| A x' - b \|_2^2 + \lambda \| x' \|_2^2 ).
\end{align*}
\end{definition}

In this paper, we study the quantum algorithms for the linear regression problem and its variations, including ridge regression and multiple regression. Quantum computing is a rapidly advancing technology, and we now have quantum computers with dramatically increasing capabilities that are on the cusp of achieving quantum advantages over classical computers. It is thus a pertinent question whether quantum computers can accelerate solving classical machine learning optimizations like linear regression.

Quantum algorithms for linear regression have been studied for a long time \cite{wan17,kp17,cd21,cgj18,sha23}. However, the majority of existing algorithms rely on quantum linear algebra techniques, which harbor noteworthy limitations. Specifically, their time complexities hinge on the condition number $\kappa$ of the input matrix. This predicament impedes a direct comparison with state-of-the-art classical methods, whose runtimes remain independent of $\kappa$. Consequently, these quantum algorithms can only guarantee acceleration over classical ones for instances featuring well-conditioned input matrices.

Overcoming this conditional dependence is an open question.  We would like quantum regression algorithms that can provably achieve speedups for any input matrix, not just ``easy'' ones.  Developing such algorithms requires departing from the quantum linear algebra framework and exploring novel techniques. In our work, we make progress in this direction by proposing quantum algorithms for linear regression, ridge regression, and multiple regression based on leverage score distribution. Our approach achieves a runtime proportional to the square root of the data dimension $n$, without the dependence on the condition number $\kappa$. This marks the first unconditional acceleration for these three regression problems in comparison to the best-known classical algorithm.

In the classical setting, it is well known that solving linear regression requires $\Omega(n)$ time \cite{cw13,nn13}. In the quantum setting, it has been known for a while there $\Omega(\sqrt{n} + d)$ times is a lower bound \cite{wan17,sha23}. Thus, in particular, we can ask the following question
\begin{center}
{\it
    Is that possible to solve linear regression in $O_d(\sqrt{n})$ time and without paying matrix-dependent parameters (e.g., $\kappa(A)$)?
}
\end{center}

In this work, provide a positive answer to this question.
 
\subsection{Our Results}

The main contribution of our work is to propose a quantum algorithm that solves linear regression in $O_d(\sqrt{n})$ time while the classical algorithm requires $\Omega_d(n)$ time. Notice that the complexity of our algorithm does not have any data-dependent parameter. For comparison, the quantum linear regression algorithms proposed by Wang \cite{wan17}, Kerenidis and Prakash \cite{kp17}, Chakraborty, Gily{\'e}n and Jeffery \cite{cgj18} have a time complexity $\poly(\log n, d, \kappa(A), \epsilon^{-1})$.  On the other hand, there exists a series of works on developing ``quantum-inspired'' algorithms for linear regression problems, which show that classical algorithms can also achieve $\log(n)$-dependence by assuming some sampling access to the input matrix. However, the time complexities of these algorithms have large polynomial dependence on some matrix parameters. In particular, Chia, Gily{\'e}n, Li, Lin, Tang, and Wang \cite{cgl22} presented a quantum-inspired algorithm that runs in  $O( ( {\| A \|_F} / { \| A \| } )^6 \kappa(A)^{28} )$ time. Further, Gily{\'e}n, Song, and Tang \cite{gst20} improved to $O( ( {\| A \|_F} / { \| A \| } )^6 \kappa(A)^{12} )$ time.  

Our result for linear regression is stated below. 
\begin{theorem}[Quantum algorithm for linear regression, informal version of Theorem~\ref{thm:linear_regression:formal}]\label{thm:linear_regression:informal}
Let $\epsilon \in (0, 1)$. Let $\omega \approx 2.37$ denote the exponent of matrix multiplication. Given a matrix $A \in \R^{n \times d}$ and $b \in \R^n$, there is a quantum algorithm that outputs $x \in \R^d$ such that
\begin{itemize}
    \item $\| A x - b \|_2 \leq (1+\epsilon)\min_{x' \in \R^d} \| A x' - b \|_2$ 
    \item it takes $\wt{O}(\sqrt{nd} / \epsilon)$ row queries to $A$ 
    \item it takes $\wt{O}(\sqrt{n} d^{1.5} / \epsilon + d^{\omega} / \epsilon )$ time, where $r$ is the row of sparsity of matrix $A$. 
    \item the success probability is 0.999
\end{itemize}
\end{theorem}

We can also improve the classical multiple linear regression algorithm from $O(nd)+ N \poly(d)$ \cite{cw13,nn13} to $\wt{O}( \sqrt{n} d^{1.5} ) + N \poly(d)$.
\begin{theorem}[Quantum algorithm for multiple regression, informal version of Theorem~\ref{thm:multiple_regression:formal}]\label{thm:multiple_regression:informal}
Let $\epsilon \in (0, 1)$. Let $\omega \approx 2.37$ denote the exponent of matrix multiplication. Given a matrix $A \in \R^{n \times d}$ with row sparsity $r$, $B \in \R^{n \times N}$, there is a quantum algorithm that outputs $X \in \R^{d \times N}$ such that
\begin{itemize}
    \item $\| A X - B\|_F \leq (1+\epsilon)\min_{X' \in \R^{d \times N}} \| A X' - B \|_F$  
    \item it takes $\wt{O}(\sqrt{nd} / \epsilon)$ row queries to $A$
    \item it takes $\wt{O}(\sqrt{n}d^{1.5} / \epsilon + d^{\omega} / \epsilon + N d^{\omega -1} / \epsilon )$ time
    \item the success probability is 0.999,
\end{itemize}
\end{theorem}

For ridge regression, the previous best ridge regression algorithm is due to \cite{acw17}, which has a running time $O(nd) + \poly( d, \mathsf{sd}_{\lambda}(A), 1/\epsilon )$. In quantum, Shao \cite{sha23} gave a quantum algorithm that has a linear dependence in $n$ in the worst case. Chen and de Wolf \cite{cd21} studied quantum algorithms for LASSO (linear regression with $\ell_1$-constraint) and ridge regressions. However, they focused on improving the $d$-dependence and only considered the regime when $n=O(\log (d)/\epsilon^2)$. In this work, we present a quantum algorithm that runs in $\wt{O}(\sqrt{n \cdot \mathsf{sd}_{\lambda}(A) }  d  ) + \poly( \mathsf{sd}_{\lambda}(A), 1/\epsilon )$ time.

\begin{theorem}[Quantum algorithm for ridge regression, informal version of Theorem~\ref{thm:ridge_regression:formal}]\label{thm:ridge_regression:informal}
Let $\epsilon \in (0, 1)$. Let $\lambda > 0$ denote a regularization parameter.  Given a matrix $A \in \R^{n \times d}$ and $b \in \R^n$. Let $\mathsf{sd}_{\lambda}(A)$ denote the statistical dimension of matrix $A$ (see Definition~\ref{def:sd}). There is a quantum algorithm that outputs $x \in \R^d$ such that
\begin{itemize}
    \item $\| A x - b \|_2^2 + \lambda \| x \|_2^2 \leq (1+\epsilon)\min_{x' \in \R^d} ( \| A x' - b \|_2^2 + \lambda \| x' \|_2^2 ) $ 
    \item it takes $\wt{O}(\sqrt{n \cdot \mathsf{sd}_{\lambda}(A) } / \epsilon)$ row queries to $A$ 
    \item it takes $\wt{O}(\sqrt{n \cdot \mathsf{sd}_{\lambda}(A) } d / \epsilon + \poly(d, \mathsf{sd}_{\lambda}(A) ,1/\epsilon) )$ time. 
    \item the success probability is 0.999
\end{itemize}
\end{theorem}

\paragraph{Roadmap.}

In Section~\ref{sec:preli}, we present the basic notations and the important mathematical definitions and properties. In Section~\ref{sec:linear_regression}, we analyze the linear regression problem and the multiple regression problem and present the formal version of our main results. In Section~\ref{sec:ridge_regression}, we analyze the ridge regression problem and present the formal version of our main result. 

\section{Preliminary}

First, we introduce the basic mathematical notations we use in this paper. Then, we introduce the important definitions and mathematical properties which support our analysis. Specifically, in Section~\ref{sub:preli:se_amp}, we introduce the definitions and properties related to the subspace embedding and approximate matrix product. In Section~\ref{sub:preli:leverage}, we formally define leverage score distribution. In Section~\ref{sub:preli:stats_dimension}, we formally define the statistical dimension. In Section~\ref{sub:preli:quantum_se}, we incorporate the quantum tools to study the properties of subspace embedding. In Section~\ref{sub:preli:approximate_matrix}, we show that the leverage score sample preserves the approximate matrix product. In Section~\ref{sub:preli:fast_matrix}, we present the running times of fast matrix multiplication.

\label{sec:preli}

\paragraph{Notations.}

First, we introduce the notations related to the sets. We define $\Z^+ : = \{1, 2, 3, \dots\}$ to be the set containing all positive integers. Let $n, d \in \Z^+$. We define $[n] := \{1, 2, 3, \dots, n\}$. We use $\R$, $\R^n$, and $\R^{n \times d}$ to denote the set containing all real numbers, all $n$-dimensional vectors with real entries, and the $n \times d$ matrices with real entries. 

Now, we introduce the notations related to vectors. Let $x \in \R^n$. For all $i \in [n]$, we let $x_i \in \R$ be the $i$-th entry of $x$. We define the $\ell_2$ norm of $x$, denoted as $\|x\|_2$, as $\sqrt{\sum_{i = 1}^n x_i^2}$.

After that, we present the notations related to the matrices. Let $A \in \R^{n \times d}$. For all $i \in [n]$, $j \in [d]$, we define $A_{i, j} \in \R$ as the entry of $A$ at the $i$-th row and $j$-th column; we define $A_{i, *} \in \R^d$ as the $i$-th row of $A$; we define $A_{*, j} \in \R^n$ as the $j$-th column of $A$. Given a vector $y \in \R^d$ satisfying $\|y\|_2 = 1$, we define the spectral norm of $A$, denoted as $\|A\|$, to be $\max_{y \in \R^d} \|Ay\|_2$. We define the Frobenius norm of $A$ as $\|A\|_F := \sqrt{\sum_{i = 1}^n \sum_{j = i}^d |A_{i, j}|^2}$. The $\ell_0$ norm of $A$, denoted as $\|A\|_0 \in \R$, is defined to be the number of nonzero entries in $A$. We use $I_d$ to represent the $d \times d$ identity matrix. We use $A^\top \in \R^{d \times n}$ to denote the transpose of the matrix $A$. $A^\dagger$ denote the pseudoinverse of $A$. Given two symmetric matrices $B, C \in \R^{n \times n}$, we use $B\preceq C$ to represent that the matrix $C-B$ is positive semidefinite (or PSD), namely for all $x \in \R^n$, we have $x^\top (C-B) x \geq 0$.

Finally, we define the notations related to functions. We use $\poly(n)$ to represent a polynomial in $n$. Let $f,g : \R \to \R$ be two functions. We use $\wt{O}(f)$ to denote $f \cdot \poly(\log f)$. We use $g(n) = O(f(n))$ to represent that there exist two positive real numbers $C$ and $x_0$ such that for all $n \geq x_0$, we have $|g(n)| \leq C \cdot f(n)$. $\arg \min_{x} f(x)$ denote the $x$ value such that $f(x)$ attains its minimum.

\subsection{Definitions of SE and AMP}
\label{sub:preli:se_amp}

In this section, we introduce key concepts that will be central to proving the guarantees of our quantum algorithms for regression. Specifically, we formally define two main concepts--subspace embedding and approximate matrix product. Subspace embedding is when multiplying by a sketching matrix approximately preserves the geometry or ``norms'' of vectors from a given subspace. Approximate matrix product means that multiplying a sketch by matrices $A$ and $B$ roughly preserves the Frobenius or spectral norm as if $A$ was directly multiplied by $B$. 

\begin{definition}[Subspace embedding, \cite{s06}]\label{def:SE}
Let $\epsilon, \delta \in (0, 1)$. Let $n > d$. Given a matrix $U \in \R^{n \times d}$ which is an orthonormal basis (i.e., $U^\top U = I_d$), we say $S\in \R^{m \times n}$ is an $\SE(\epsilon,\delta,n,d)$ subspace embedding for $U$ if 
\begin{align*} 
    \| S U x \|_2^2 = (1 \pm \epsilon) \| U x \|_2^2,
\end{align*}
holds with probability $1-\delta$.

This is equivalent to
\begin{align*}
    \| U^\top S^\top S U - U^\top U \| \leq \epsilon
\end{align*}
\end{definition}

\begin{remark}
In general, if $S$ is not depending on $U$ we call it is oblivious subspace embedding. In the most places of this paper, our $S$ is depending on $U$. That's why we don't use ``oblivious'' in the definition like other papers \cite{syyz23_linf}.
\end{remark}

\begin{definition}[Frobenius Norm Approximate Matrix Product, \cite{w14}]\label{def:FAMP}
Let $\epsilon, \delta \in (0, 1)$. We say $S \in \R^{m \times n}$ is $\FAMP(\epsilon, \delta, n,d)$ Approximate Matrix Product for $A\in \R^{n \times d}$ if for any $B\in \R^{n \times N}$ we have
\begin{align*}
   \| A^\top S^\top S B - A^\top B \|_F^2 \leq \epsilon^2 \cdot \| A \|_F^2 \cdot \| B \|_F^2
\end{align*}
holds with probability $1-\delta$.
\end{definition}
\begin{remark}
Here matrix $B$ has to have the same number of rows as $A$. However, $B$ is not necessarily to have the same number of columns as $A$. 
\end{remark}

\begin{definition}[Spectral Norm Approximate Matrix Product, see Theorem 17 in \cite{acw17}]\label{def:SAMP}
Let $\epsilon, \delta \in (0, 1)$. We say $S \in \R^{m \times n}$ is $\SAMP(\epsilon, \delta, n,d)$ Approximate Matrix Product for $A\in \R^{n \times d}$ if for any $B\in \R^{n \times N}$ we have
\begin{align*}
   \| A^\top S^\top S B - A^\top B \| \leq \epsilon \cdot \| A \| \cdot \| B \|
\end{align*}
holds with probability $1-\delta$.
\end{definition}
\begin{remark}
Here matrix $B$ has to have the same number of rows as $A$. However, $B$ is not necessarily to have the same number of columns as $A$. 
\end{remark}

\begin{claim}\label{cla:equiv}
If the following conditions hold
\begin{itemize}
\item Let $ A \in \R^{n \times d}$. 
\item Let $U$ denote the orthonormal basis of $A$.
\item Let $D$ denote a diagonal matrix such that $\| D A x \|_2^2 = (1\pm \epsilon) \| A x \|_2^2$ for all $x$.
\end{itemize}
Then, we have 
\begin{align*}
    \| D U x \|_2^2 = (1\pm \epsilon) \| U x \|_2^2
\end{align*}
\end{claim}
\begin{proof}
Let $R \in \R^{d \times d}$ denote the QR factorization of $A$. Then we have
\begin{align*}
    A = U R
\end{align*} 

From $\| DAx \|_2^2 = (1\pm \epsilon) \| A x \|_2^2$, we know that
\begin{align*}
    \| D U R x \|_2^2 =  (1\pm \epsilon) \| U R x \|_2^2, \forall x
\end{align*}
Since $R$ is full rank, then we can replace $Rx$ by $y$ to obtain
\begin{align*}
    \| D U y \|_2^2 =  (1\pm \epsilon) \| U y \|_2^2,  \forall y.
\end{align*}
\end{proof}

\subsection{Leverage Score Distribution}
\label{sub:preli:leverage}

We introduce leverage score (see Definition~\ref{def:leverage_score}) and leverage score distribution (see Definition~\ref{def:ls_distribution}), which are well-known concepts in numerical linear algebra. We provide definitions that quantify the leverage score of a matrix row as its squared Euclidean norm under an orthonormal transformation of the matrix. Additionally, we define a leverage score distribution as a probability distribution that samples rows with probabilities proportional to these row leverage scores. Intuitively, leverage scores control how much influence each row vector has in spanning the column space.

\begin{definition}[Leverage Score, see Definition B.28 in \cite{swz19} as an example]\label{def:leverage_score}
Given a matrix $A \in \R^{n \times d}$, we let $U \in \R^{n \times d}$ denote the orthonormal basis of $A$. We define $\sigma_i := \| U_{i,*} \|_2^2$ for each $i \in [n]$. We say $\sigma \in \R^n$ is the leverage score of $A \in \R^{n \times d}$.  
\end{definition}

\begin{fact}
It is well known that $\sum_{i=1}^n \sigma_i = d$.
\end{fact}

\begin{definition}[$D \sim \mathsf{LS}(A)$, see Definition B.29 in \cite{swz19} as an example]\label{def:ls_distribution}
 Let $c > 1$ denote some universal constant. 

 For each $i \in [n]$, we define $p_i := c \cdot \sigma_i / d$.  

 Let $q \in \R^n$ be the vector that $q_i \geq p_i$.

 Let $m$ denote the sparsity of diagonal matrix $D \in \R^{n \times n}$.
 
 We say a diagonal matrix $D$ is a sampling and rescaling matrix according to leverage score of $A$ if for each $i \in [n]$, $D_{i,i} = \frac{1}{\sqrt{m q_i} }$ with probability $q_i$ and $0$ otherwise. (Note that each $i$ is picked independently and with replacement) We use notion $D \sim \mathsf{LS}(A)$ to denote that.
\end{definition}

\subsection{Statistical Dimension}
\label{sub:preli:stats_dimension}

In this section, we formally define the statistical dimension. In addition to leverage scores capturing the geometric influence of matrix rows, another related notion that will play an important role is the statistical dimension. While leverage scores are row-specific concepts, statistical dimension provides an aggregate measure of the complexity of a matrix that governs the sample size needed to effectively sketch it.

\begin{definition}[Statistical dimension, see Definition 1 in \cite{acw17} as an example]\label{def:sd}
For real value $\lambda \geq 0$ and rank-$d$ matrix $A \in \R^{n \times d}$ with singular values $\sigma_i(A)$, the quantity $\mathsf{sd}_{\lambda}(A):= \sum_{i=1}^d \frac{1}{1+ \lambda/\sigma_i(A)^2}$ is the statistical dimension of the ridge regression problem with regularizing weight $\lambda$.

\end{definition}

\subsection{Quantum Tools for Subspace Embedding}
\label{sub:preli:quantum_se}

In this section, our purpose is to bridge the gap between classical theory and quantum techniques: we present a quantum tool we use for designing a fast algorithm based on leverage score sampling. This tool was recently developed by Apers and Gribling \cite{ag23}. 
In particular, we state two Lemmas analyzing efficient quantum routines for leverage score estimation and sampling, respectively. The first allows approximating row leverage scores (see Definition~\ref{def:leverage_score}) with queries to the input matrix scaling as the square root of the dimension. The second one constructs a sampling matrix from these estimated scores that serves as a subspace embedding (see Definition~\ref{def:SE}).

\begin{lemma}[Theorem 3.2 in \cite{ag23}]\label{lem:ag23_theorem_3.1}
Consider query access to matrix $A \in \R^{n \times d}$ with row sparsity $r$. For any $\epsilon_0 \in (0,1)$, there is a quantum algorithm that, provides query access to estimate $\wt{\sigma}_i$ for any $i \in [n]$ satisfying $\wt{\sigma}_i = (1\pm \epsilon_0 ) \sigma(A)_i$
\begin{itemize}
\item It makes $\wt{O}(\sqrt{nd} / \epsilon_0 )$ row queries to $A$.
\item It takes $\wt{O}( r \sqrt{nd} / \epsilon_0 + d^{\omega} / \epsilon_0^2 + d^2 / \epsilon_0^4 )$ time.
\item The success probability $0.999$
\item Cost per estimate $\wt{\sigma}_i$: one row query to $A$ and $\wt{O}(r/\epsilon_0^2)$ time
\end{itemize}
\end{lemma}

\begin{lemma}\label{lem:ag23_tool}
Consider query access to matrix $A \in \R^{n \times d}$ with row sparsity $r$. Let $U$ denote the orthonormal basis of $A$. For any $\epsilon \in (0,1)$, there is a quantum algorithm that, returns a diagonal matrix $D \in \R^{n \times n}$ such that
\begin{itemize}
\item $\| D \|_0 = O( \epsilon^{-2} d \log d )$
\item $\| D U x \|_2^2 =(1\pm \epsilon) \| U x \|_2^2$ (subspace embedding)
\item $D \sim \mathsf{LS}(A)$ (see Definition~\ref{def:ls_distribution})
\item It makes $\wt{O}(\sqrt{nd} / \epsilon )$ row queries to $A$.
\item It takes $\wt{O}( r \sqrt{nd} / \epsilon + d^{\omega} )$ time.
\item The success probability $0.999$
\end{itemize}
\end{lemma}
\begin{proof}
To do the leverage score sampling, we only need to set $\epsilon_0 = 0.1$ to be constant in Lemma~\ref{lem:ag23_theorem_3.1}.

Using classical correctness, we know that if the sampling size is $O(\epsilon^{-2} d \log d)$, then we will get subspace embedding (see Definition~\ref{def:SE}).

Using quantum sampling lemma, we know that sampling $\| D\|_0$ rows from an $n$ rows of $A$ requires $\sqrt{n \| D \|_0}$ row queries to $A$.

Using Lemma~\ref{lem:ag23_theorem_3.1} it takes 
\begin{align*}
    \wt{O}( r \sqrt{n \| D \|_0} + d^\omega )
\end{align*}
Thus, we complete the proof.
\end{proof}

\subsection{Leverage Score Sample Preserves Approximate Matrix Product}
\label{sub:preli:approximate_matrix}

In this section, we show that the leverage score sample preserves the approximate matrix product. Specifically, we analyze the Lemma showing that sampling rows of a matrix $A$ proportionally to leverage scores (see Definition~\ref{def:leverage_score}) generates a sketching matrix that approximates the product between $A$ and any other matrix $B$ with respect to the Frobenius norm.

\begin{lemma}[Lemma C.29 in \cite{swz19}]
If the following conditions hold
\begin{itemize}
    \item Let $A \in \R^{n \times d}$
    \item Let $B \in \R^{n \times N}$
    \item Let $\epsilon \in (0,1)$
    \item Let $S \sim \mathsf{LS}(A)$ (see Definition~\ref{def:ls_distribution})
    \item Let $\| S \|_0 = O(1/\epsilon^2)$ 
\end{itemize}
Then, for any fixed matrix $B$, we have
\begin{itemize}
    \item $\| A^\top S^\top S B - A^\top B \|_F^2 \leq \epsilon^2 \| A \|_F^2 \| B \|_F^2$
    \item The success probability is $0.999$
\end{itemize}
\end{lemma}

\subsection{Fast Matrix Multiplication}
\label{sub:preli:fast_matrix}

In this section, we present the running time of the fast matrix multiplication. We define the variable $\Tmat$ to represent the time cost of multiplying two matrices of designated dimensions. Further, we use $\omega$ to refer to the matrix multiplication exponent governing the asymptotic scaling of these runtimes \cite{w12,lg14,aw21,dwz23,lg23,wxxz23}. We also introduce useful facts about manipulating the matrix multiplication time function. These rules will assist with analyzing the runtimes of operations like computing $(SA)^\dagger SB$ which occur inside our regression analysis (see Lemma~\ref{lem:linear_running_time}).

\begin{definition}
Given two matrices $a \times b$ size and $b \times c$, we use the $\Tmat(a,b,c)$ to denote the time of multiplying $a \times b$ matrix with another $b \times c$.
\end{definition}

We use $\omega$ to denote the number that $\Tmat(n,n,n) = n^{\omega}$.

\begin{fact}\label{fac:mm_swap}
Given three positive integers, we have
\begin{align*}
    \Tmat(a,b,c) = O(\Tmat(a,c,b)) = O(\Tmat(b,a,c)) = O(\Tmat(b,c,a)) = O(\Tmat(c,a,b)) = O(\Tmat(c,b,a) )
\end{align*}
\end{fact}

\begin{fact}\label{fac:mm_divide}
Given $a,b,c,d$ are positive integers. Then we have
\begin{itemize}
\item {\bf Part 1.}
\begin{align*}
    \Tmat(a,b,c) = O( d\cdot \Tmat(a/d,b,c) )
\end{align*}
\item {\bf Part 2.}
\begin{align*}
    \Tmat(a,b,c) = O( d\cdot \Tmat(a,b/d,c) )
\end{align*}
\item {\bf Part 3.}
\begin{align*}
    \Tmat(a,b,c) = O( d\cdot \Tmat(a,b,c/d) )
\end{align*}
\end{itemize}
\end{fact}

\section{Multiple Regression and Linear Regression}
\label{sec:linear_regression}

In this section, we focus on proving the result for multiple regression. It directly implies the result for linear regression.

In Section~\ref{sub:linear_regression:ls_se_amp}, we show that the leverage score distribution may imply the subspace embedding and approximate matrix product. In Section~\ref{lem:se_amp_reg}, we show that by using the subspace embedding and approximate matrix product, we can get the multiple regression. In Section~\ref{sub:linear_regression:running_time}, we analyze the running time for each of the matrices $SA$, $SB$, $(SA)^\dagger$, and $(SA)^\dagger \cdot (SB)$. In Section~\ref{sub:linear_regression:main}, we combine the important properties of this section to form the formal version of our result for multiple regression, and based on that, we take $N = 1$ to form the formal version of our result for linear regression.

\subsection{LS Implies SE and AMP}
\label{sub:linear_regression:ls_se_amp}

In this section, we present a tool from Song, Woodruff, and Zhong \cite{swz19} showing that if $S \sim \mathsf{LS}(A)$, a leverage score distribution, then $S$ is a subspace embedding (see Definition~\ref{def:SE}) and satisfies the definition of Frobenius norm approximate matrix product (see Definition~\ref{def:FAMP}). The purpose is to rigorously justify why sampling from leverage scores enables dimensionality reduction for regression problems. By showing that the sampled matrix retains the structure of the original matrix, we lay the groundwork to prove that solving regression on the smaller sampled matrix yields an approximate solution for the full regression problem. This then sets up the development in later sections showing how this sampling-based reduction leads to faster quantum algorithms.

\begin{lemma}[Corollary C.30 in \cite{swz19}]\label{lem:ls_se_amp}
If the following conditions hold
\begin{itemize}
    \item Given $A \in \R^{n \times d}$
    \item Let $U$ denote the orthonormal basis of $A$
    \item Let $S \sim \mathsf{LS}(A)$ (see Definition~\ref{def:ls_distribution})
    \item Let $\| S \|_0 = m$
\end{itemize}
Then, we have
\begin{itemize}
    \item {\bf Part 1.} If $m = O(d \log d)$, then $S$ is a $\SE(1/2,0.99, n,d)$ subspace embedding (see Definition~\ref{def:SE}) for $U$
    \item {\bf Part 2.} If $m = O(d/\epsilon)$, then $S$ satisfies $\FAMP(\sqrt{\epsilon/d},0.99,n,d)$ (see Definition~\ref{def:FAMP}) for $U$
\end{itemize}
\end{lemma}

\subsection{From SE and AMP to Regression}
\label{sub:linear_regression:se_amp_reg}

In this section, we present another tool from \cite{swz19} showing that if $S$ is a subspace embedding (see Definition~\ref{def:SE}) and satisfies the definition of Frobenius norm approximate matrix product (see Definition~\ref{def:FAMP}), then the multiple regression is satisfied. Therefore, solving the sketched regression problem on the smaller matrix $SA$ yields solutions that generalize to approximate the full regression problem on $A$.

\begin{lemma}[Lemma C.31 in \cite{swz19}]\label{lem:se_amp_reg}
If the following conditions hold
\begin{itemize}
    \item Let $A \in \R^{n \times d}$
    \item Let $B \in \R^{n \times N}$
    \item Let $S \sim \mathsf{LS}(A)$ (see Definition~\ref{def:ls_distribution}). 
    \item Let $X^* = \arg\min_{X} \| A X - B \|_F^2$
    \item Let $X' = \arg\min_{X} \| SA X - S B \|_F^2$ 
    \item Let $U$ denote an orthonormal basis for $A$
    \item Suppose $S$ is $\SE(1/2, 0.99,n,d)$ (see Definition~\ref{def:SE}) for $U$
    \item Suppose $S$ is $\FAMP(\sqrt{\epsilon/d},0.99,n,d)$ (see Definition~\ref{def:FAMP}) for $U$
\end{itemize}
Then, we have
\begin{align*}
    \| A X' - B \|_F^2 \leq (1+\epsilon) \| A X^* - B \|_F^2
\end{align*}
\end{lemma}

\subsection{Computing the Running Time}
\label{sub:linear_regression:running_time}

In this section, we state and prove a lemma bounding the running time for using leverage score sampling (see Definition~\ref{def:ls_distribution}), then forming the sketched matrices $SA$ and $SB$, computing the pseudoinverse of $SA$, and multiplying this pseudoinverse by $SB$ to obtain the sketched solution. Each of these pieces is analyzed in terms of the input dimension $n$, $d$, and the accuracy parameter $\epsilon$. The purpose of this section is to complement the correctness guarantees from Section~\ref{sub:linear_regression:ls_se_amp} and Section~\ref{sub:linear_regression:se_amp_reg} by quantifying the computational efficiency of the sampling reduction process. This runtime analysis will be essential in establishing the time complexities of our final quantum regression algorithms.

\begin{lemma}\label{lem:linear_running_time}
If the following conditions hold
\begin{itemize}
    \item Let $A \in \R^{n \times d}$
    \item Let $\epsilon \in (0, 0.1)$
    \item Let $B \in \R^{n \times N}$
    \item Let $\omega \approx 2.37$
    \item Let $S$ denote a diagonal matrix that $\| S \|_0 = O(d\log d + d/\epsilon)$
    \item $X' = (SA )^\dagger SB$
\end{itemize}
Then, we have we can compute $X' \in \R^{d \times N}$ in $\wt{O}( d^{\omega}/\epsilon + N d^{\omega-1} / \epsilon )$ 
\end{lemma}
\begin{proof}
The proof directly follows from computing the time of $SA$, $SB$, $(SA)^\dagger$, and $(SA)^\dagger \cdot (SB)$.

Before computing the running time, let us recall the definition
\begin{itemize}
    \item $S$ is an $n \times n$ size diagonal entries which only has $\wt{O}(d/\epsilon)$ non-entries on diagonal
    \item $A$ has size $n \times d$
\end{itemize}

Here are the computation costs:
\begin{itemize}
    \item Computing $\wt{O}(d/\epsilon) \times d$ size matrix $SA$ takes $\wt{O}(d^2/\epsilon)$ 
    \item Computing $\wt{O}(d/\epsilon) \times N$ size matrix $SB$ takes $\wt{O}(N d / \epsilon)$ time
    \item Computing $d \times \wt{O}(d/\epsilon)$ size matrix $(SA)^\dagger$ takes $\wt{O}( d^{\omega}/\epsilon )$ time 
    \item Computing $d \times N$ size matrix $(SA)^\dagger (SB)$ takes $\Tmat( d, \wt{O}(d/\epsilon) , N ) = \wt{O}(N d^{\omega-1} / \epsilon)$ time (due to Fact~\ref{fac:mm_divide}).
\end{itemize}
Thus, we complete the proof.
\end{proof}

\subsection{Main Result}
\label{sub:linear_regression:main}

At this point, we have developed all the theoretical concepts required to obtain faster quantum algorithms for regression problems based on a sampling reduction approach. Thus, in this section, we present our main results for the multiple regression and the linear regression. First, we incorporate the mathematical properties developed earlier to present our result for the multiple regression.

\begin{theorem}[Multiple Regression, our result, formal version of Theorem~\ref{thm:multiple_regression:informal}]\label{thm:multiple_regression:formal}
Let $\epsilon \in (0, 1)$. Let $\omega \approx 2.37$ denote the exponent of matrix multiplication. Given a matrix $A \in \R^{n \times d}$ with row sparsity $r$, $B \in \R^{n \times N}$, there is a quantum algorithm that outputs $X \in \R^{d \times N}$ such that
\begin{itemize}
    \item $\| A X - B\|_F \leq (1+\epsilon)\min_{X' \in \R^{d \times N}} \| A X' - B \|_F$  
    \item it takes $\wt{O}(\sqrt{nd} / \epsilon)$ row queries to $A$
    \item it takes $\wt{O}(r\sqrt{nd} / \epsilon + d^{\omega} / \epsilon + N d^{\omega -1} / \epsilon )$ time, where $r$ is the row of sparsity of matrix $A$. 
    \item the success probability is 0.999,
\end{itemize}
\end{theorem}
\begin{proof}

It follows from combining Lemma~\ref{lem:ag23_tool}, Lemma~\ref{lem:ls_se_amp}, Lemma~\ref{lem:se_amp_reg}, and Lemma~\ref{lem:linear_running_time}.

Note that by combining Lemma~\ref{lem:ls_se_amp} and Lemma~\ref{lem:se_amp_reg}, we can have
\begin{align*}
    \| A X' - B \|_F^2 \leq (1+\epsilon) \| A X^* - B \|_F^2.
\end{align*}

Lemma~\ref{lem:ag23_tool} and Lemma~\ref{lem:linear_running_time} give us the running time.
\end{proof}

Then, we present our result for the linear regression, where we treat this as the multiple regression for $N = 1$.

\begin{theorem}[Linear Regression, our result, formal version of Theorem~\ref{thm:linear_regression:informal}]\label{thm:linear_regression:formal}

Let $\epsilon \in (0, 1)$. Let $\omega \approx 2.37$ denote the exponent of matrix multiplication. Given a matrix $A \in \R^{n \times d}$ and $b \in \R^n$, there is a quantum algorithm that outputs $x \in \R^d$ such that
\begin{itemize}
    \item $\| A x - b \|_2 \leq (1+\epsilon)\min_{x' \in \R^d} \| A x' - b \|_2$ 
    \item it takes $\wt{O}(\sqrt{nd} / \epsilon)$ row queries to $A$ 
    \item it takes $\wt{O}(\sqrt{n} d^{1.5} / \epsilon + d^{\omega} / \epsilon )$ time, where $r$ is the row of sparsity of matrix $A$. 
    \item the success probability is 0.999
\end{itemize}
    
\end{theorem}

\begin{proof}

Let $x : = X \in \R^{d \times N}$ when $N = 1$.

Let $b : = B \in \R^{n \times N}$ when $N = 1$.

Then, by Theorem~\ref{thm:multiple_regression:formal}, we have
\begin{align*}
    \| A x - b \|_2 \leq (1+\epsilon)\min_{x' \in \R^d} \| A x' - b \|_2,
\end{align*}
which takes
\begin{align*}
    \wt{O}(r\sqrt{nd} / \epsilon + d^{\omega} / \epsilon + d^{\omega -1} / \epsilon ) = \wt{O}(\sqrt{n} d^{1.5} / \epsilon + d^{\omega} / \epsilon )
\end{align*}
time.
\end{proof}

\section{Ridge Regression}
\label{sec:ridge_regression}

In Section~\ref{sub:ridge_regression:orthonormal}, we present a property of the orthonormal basis for the ridge matrix. In Section~\ref{sub:ridge_regression:sampling}, we introduce a sampling oracle related to $U_1$. In Section~\ref{sub:ridge_regression:ls_se_samp}, we present the property of the leverage score distribution that for a matrix $S$ sampled from it, $S$ is a subspace embedding (Definition~\ref{def:SE}) and satisfy the definition of the spectral norm approximate matrix product (Definition~\ref{def:SAMP}). In Section~\ref{sub:ridge_regression:sketch}, we present the property of the matrix if it is a subspace embedding and satisfy the definition of the spectral norm approximate matrix product. In Section~\ref{sub:ridge_regression:main}, we present our main result for the ridge regression.

\subsection{Property of Orthonormal Basis for Ridge Matrix}
\label{sub:ridge_regression:orthonormal}

In order to develop our quantum ridge regression algorithm, we require an efficient way to reduce the ridge regression problem on the original matrix $A$ to a regularized regression problem on a much smaller sampled matrix. Therefore, in this section, we present the property of the orthonormal basis for the ridge matrix. The guarantee relating the original ridge regression problem on $A$ and the reduced regression problem on $U_1$, which comprises the first $n$ rows of the orthonormal basis of the augmented matrix $\begin{bmatrix} A \\ \sqrt{\lambda} I_d \end{bmatrix}$, is formally presented in the following claim. This claim provides an explicit formula for the squared Frobenius norm of $U_1$ in terms of the statistical dimension of the ridge problem. It also bounds the spectral norm of $U_1$, which will be useful for ensuring subspace embedding properties (see Definition~\ref{def:SE}) when we subsample the ridge leverage scores of $U_1$.

\begin{claim}[\cite{htf09}, Lemma 12 in \cite{acw17}]\label{cla:12_in_acw17}
If the following conditions hold
\begin{itemize}
    \item Given matrix $A \in \R^{n \times d}$
    \item Let $U \in \R^{n \times d}$ denote the orthonormal basis of $A$
     \item Let $U_1 \in \R^{ n \times d}$ comprise the first $n$ rows of orthonormal basis of $\begin{bmatrix} A \\ \sqrt{\lambda} I_d \end{bmatrix}$
     \item For each $i \in [d]$, let $\sigma_i(A)$ denote the singular value of matrix $A$
\end{itemize}
Then we have
\begin{itemize}
    \item Part 1. $\| U \|_F^2 = d$
    \item Part 2. $\| U \| = 1$
    \item Part 3. $\| U_1 \|_F^2 = \sum_{i=1}^d \frac{1}{1+ \lambda/ \sigma_i(A)^2} = \mathsf{sd}_{\lambda}(A)$
    \item Part 4. $\| U_1 \| = \frac{1}{ \sqrt{1 + \lambda / \sigma_1^2 } }$
\end{itemize}
\end{claim}
\begin{proof}
For $\| U \|_F^2 = d$, it trivially follows from the definition of the orthonormal basis.

We consider the SVD\footnote{Here we use a different shape of SVD, which is not as usual $\Sigma \in \R^{d \times d}$} of 
\begin{align}\label{eq:svd_A}
    A = U \Sigma V^\top,
\end{align}
where $U \in \R^{n \times n}$, $\Sigma \in \R^{n \times d}$ and $V \in \R^{d \times d}$.

We define 
\begin{align}\label{eq:D}
    D:= ( \Sigma^{\top} \Sigma + \lambda I_d )^{-1/2}.
\end{align}
We define
\begin{align}\label{eq:A}
\wh{A} :=
\begin{bmatrix}
U \Sigma D \\
V \sqrt{\lambda} D 
\end{bmatrix}
\end{align}
Then we have
\begin{align*}
    \wh{A}^\top \wh{A} = I_d
\end{align*}

For any $x$, we define $y$
\begin{align}\label{eq:y}
 y:= D^{-1} V^\top x
\end{align}

Then, we have 
\begin{align*}
    \wh{A} y 
    = & ~ \begin{bmatrix}
U \Sigma D \\
V \sqrt{\lambda} D 
\end{bmatrix} D^{-1} V^\top x\\
= & ~ \begin{bmatrix}
U \Sigma D D^{-1} V^\top \\
V \sqrt{\lambda} D D^{-1} V^\top 
\end{bmatrix} x\\
= & ~ \begin{bmatrix}
U \Sigma V^\top \\
\sqrt{\lambda} V V^\top 
\end{bmatrix} x\\
    = & ~ \begin{bmatrix} A \\ \sqrt{\lambda} I_d \end{bmatrix} x,
\end{align*}
where the first step follows from the definition of $\wh{A}$ (see Eq.~\eqref{eq:A}) and $ y $ (see Eq.~\eqref{eq:y}), the second step follows from simple algebra, the third step follows from the fact that $DD^{-1}$ is the identity matrix, and the last step follows from the SVD of $A$ (see Eq.~\eqref{eq:svd_A}) and the fact that $V$ is orthogonal.

Finally, we can show
\begin{align*}
    \| U_1 \|_F^2 = & ~ \| U \Sigma D \|_F^2 \\
    = & ~ \| \Sigma D \|_F^2 \\ 
    = & ~ \sum_{i=1}^d \frac{1}{ 1 + \lambda / \sigma_i(A)^2 },
\end{align*}
where the first step follows from the Lemma statement, the second step follows from $U$ is an $n \times n$ orthonormal basis, and the last step follows from the definition of statistical dimension (see Definition~\ref{def:sd}).
\end{proof}

\subsection{Sampling Oracle related to \texorpdfstring{$U_1$}{}}

In this section, we present a sampling oracle related to $U_1$. Specifically, given query access to an $n \times d$ matrix $A$, we show how to sample rows from $A$ according to the ridge leverage score distribution of $U_1$ in input sparsity time. This sampling oracle serves as a crucial subroutine in our quantum ridge regression algorithm, enabling us to reduce solving the ridge regression problem on $A$ to solving a sampled regular regression problem on a much smaller sampled matrix.

\label{sub:ridge_regression:sampling}

\begin{lemma}[Sampling oracle, a ridge regression version of Lemma~\ref{lem:ag23_tool}]\label{lem:ag23_tool_sd}
Consider query access to matrix $A \in \R^{n \times d}$ with row sparsity $r$. Let $U_1$ denote the comprise of first $n$ row of orthonormal basis of $\wt{A} = \begin{bmatrix} A \\ \sqrt{\lambda} I_d \end{bmatrix}$. For any $\epsilon \in (0,1)$, there is a quantum algorithm that, returns a diagonal matrix $D \in \R^{n \times n}$ such that
\begin{itemize}
\item  Let $\| D \|_0$ denote the sparsity of $D$
\item $D \sim \mathsf{LS}( \wt{A}_{ 1:n })$ (see Definition~\ref{def:ls_distribution})
    \begin{itemize}
    \item $\wt{A}_{1:n}$ is the distribution with respect to $U_1$
    \end{itemize}
\item It makes $\wt{O}(\sqrt{n \cdot \| D \|_0 } )$ row queries to $A$.
\item It takes $\wt{O}( r \sqrt{n \| D \|_0 }  + \poly(d, \| D \|_0) )$ time.
\item The success probability $0.999$
\end{itemize}
\end{lemma}
\begin{proof}
The proof is similar to Lemma~\ref{lem:ag23_tool}. The major difference is, in Lemma~\ref{lem:ag23_tool}, we estimate a distribution with $n$ scores, and take samples from it.

Here, we estimate a distribution $n+d$ scores (with respect to $\wt{A} \in \R^{ (n+d) \times d}$), but we only take samples from first $n$ scores (with respect to $U_1 \in \R^{n \times d}$). Since the summation of first $n$ scores is fixed and can be explicitly computed (see Claim~\ref{cla:12_in_acw17} for computation of $\| U_1 \|_F^2$). Thus, our sampling is correct.
\end{proof}

\subsection{From LS to SE and SAMP}
\label{sub:ridge_regression:ls_se_samp}

With the sampling oracle and the structural ridge regression properties established in the previous sections, we now have the key ingredients to show the reduction from ridge regression on $A$ to subsampled regular regression. In this section, we present a tool from Avron, Clarkson and Woodruff \cite{acw17}, which shows that if $S$ is sampled from the leverage score distribution, then it is a subspace embedding and spectral norm approximate matrix product. The formal guarantees showing that the ridge leverage score sampler $S$ satisfies the desired properties with respect to $U_1$ are presented in the following Lemma. This lemma helps establish the validity of our approach to subsample the ridge regression problem and solve the smaller sketched problem. In later sections, we show how solving this sketched regression problem leads to fast quantum algorithms for ridge regression.

\begin{lemma}[Theorem 16 in \cite{acw17}]\label{lem:16_in_acw17}
If the following conditions hold
\begin{itemize}
    \item Given matrix $A \in \R^{n \times d}$
    \item Let $U_1 \in \R^{ n \times d}$ comprise the first $n$ rows of the orthonormal basis of $\begin{bmatrix} A \\ \sqrt{\lambda} I_d \end{bmatrix}$
    \item Let $S\sim \mathsf{LS}(A)$
    \item Let $ \| S \|_0 = \wt{O}(\epsilon^{-1} \mathsf{sd}_{\lambda} (A) )$. 
\end{itemize}
Then we have
\begin{itemize}
    \item $S$ is $\mathsf{SE}(1/2,0.99,n,d)$ for $U_1$
    \item $S$ is $\mathsf{SAMP}(\sqrt{\epsilon}, 0.99, n,d)$ for $U_1$
\end{itemize}
\end{lemma}

\subsection{Guarantee of Sketched Solution}
\label{sub:ridge_regression:sketch}
What remains is to formally argue that solving the sketched regression problem on the sampled matrix $SU_1$ yields a good approximation to the original ridge regression problem on $A$. 
In this section, we fill this missing step by showing that the guarantees provided by subspace embedding and spectral norm approximate matrix product on $U_1$ imply that the ridge regression objective is well-preserved between the original problem and the sketched problem.

In particular, in the following Lemma, we prove that if $S$ acts as a subspace embedding and spectral norm approximate matrix product for $U_1$, then the ridge regression solution on $SU_1$ yields an approximate solution for the ridge regression problem on $A$. Combining this with the sampling results from Section~\ref{sub:ridge_regression:ls_se_samp} completes the full reduction argument, demonstrating that we can subsample the ridge leverage scores of $U_1$ to reduce ridge regression on $A$ to a much smaller sketched regression problem.

\begin{lemma}[Lemma 11 in \cite{acw17}]\label{lem:11_in_acw17}
If the following conditions hold
\begin{itemize}
    \item Given matrix $A \in \R^{n \times d}$
    \item Let $U_1 \in \R^{ n  \times d}$ comprise the first $n$ rows of orthonormal basis of $\begin{bmatrix} A \\ \sqrt{\lambda} I_d \end{bmatrix}$
    \item Suppose $S$ is $\mathsf{SE}(1/2,0.99,n,d)$ for $U_1$
    \item Suppose $S$ is $\mathsf{SAMP}(\sqrt{\epsilon}, 0.99, n,d)$ for $U_1$
\end{itemize}
Then, we have 
\begin{align*}
\| Ax' -b \|_2^2 + \lambda \| x'\|_2^2 \leq (1+\epsilon) \cdot ( \| A x^* - b \|_2^2 + \lambda \| x^* \|_2^2 )
\end{align*}
\end{lemma}

\subsection{Main Result}
\label{sub:ridge_regression:main}

In this section, we present our main result for the ridge regression, bringing together the sampling oracle and the reduction arguments based on subspace embedding and approximate matrix multiplication. 

\begin{theorem}[Ridge Regression, our result, formal version of Theorem~\ref{thm:ridge_regression:informal}]\label{thm:ridge_regression:formal}
Let $\epsilon \in (0, 1)$. Let $\lambda > 0$ denote a regularization parameter.  Given a matrix $A \in \R^{n \times d}$ and $b \in \R^n$. Let $\mathsf{sd}_{\lambda}(A)$ denote the statistical dimension of matrix $A$ (see Definition~\ref{def:sd}). There is a quantum algorithm that outputs $x \in \R^d$ such that
\begin{itemize}
    \item $\| A x - b \|_2^2 + \lambda \| x \|_2^2 \leq (1+\epsilon)\min_{x' \in \R^d} ( \| A x' - b \|_2^2 + \lambda \| x' \|_2^2 ) $ 
    \item it takes $\wt{O}(\sqrt{n \cdot \mathsf{sd}_{\lambda}(A) } / \epsilon)$ row queries to $A$ 
    \item it takes $\wt{O}(\sqrt{n \cdot  \mathsf{sd}_{\lambda}(A) } \cdot d  / \epsilon + \poly(d, \mathsf{sd}_{\lambda}(A) ,1/\epsilon) )$ time. 
    \item the success probability is 0.999
\end{itemize}
\end{theorem}
\begin{proof}
It follows from combining Lemma~\ref{lem:ag23_tool}, Lemma~\ref{lem:ag23_tool_sd}, Claim~\ref{cla:12_in_acw17}, Lemma~\ref{lem:16_in_acw17}, and Lemma~\ref{lem:11_in_acw17}.  
\end{proof}

\ifdefined\isarxiv
\bibliographystyle{alpha}
\bibliography{ref}
\else
\bibliography{ref}
\bibliographystyle{alpha}

\fi

\newpage
\onecolumn
\appendix




\end{document}